\documentclass[runningheads]{llncs}

\usepackage{eccv}

\usepackage{eccvabbrv}

\usepackage{graphicx}
\usepackage{booktabs}

\usepackage[accsupp]{axessibility}  

\usepackage[pagebackref,breaklinks,colorlinks,citecolor=eccvblue]{hyperref}

\usepackage{orcidlink}

\usepackage{amsmath,amsfonts,bm}

\def\eqref#1{equation~\ref{#1}}

\def\floor#1{\lfloor #1 \rfloor}
\def\1{\bm{1}}

\DeclareMathAlphabet{\mathsfit}{\encodingdefault}{\sfdefault}{m}{sl}
\SetMathAlphabet{\mathsfit}{bold}{\encodingdefault}{\sfdefault}{bx}{n}

\usepackage[dvipsnames]{xcolor}

\usepackage{multirow}
\usepackage{algorithm, algpseudocode}
\usepackage{graphicx}
\usepackage{pgfplots}
\usepackage{comment}
\usepackage{subcaption}
\usepackage{listings}

\author{    Jonathan Rosenthal, Shanchao Liang, Kevin Zhang, \& Lin Tan}

\institute{Purdue University, West Lafayette, IN 47906, USA
\email{\{rosenth0,liang422,zhan4196,lintan\}@purdue.edu}}

\usepackage{adjustbox}

\renewcommand{\Comment}[1]{}

\definecolor{ultrapink}{rgb}{1.0, 0.44, 1.0}
\definecolor{neongreen}{rgb}{0.6, 0.9, 0}
\definecolor{darkgreen}{rgb}{0, .65, 0}
\definecolor{orange}{RGB}{255,165,0}

\newcommand{\todoc}[2]{{\textcolor{#1}{\textbf{#2}}}}

\definecolor{light-gray}{gray}{0.7}

\renewcommand{\todoc}[2]{\relax}

\newcommand{\ClassBalanced}{{Class-Balanced Difficulty-Weighted Replay}\xspace}
\newcommand{\Classbalanced}{{Class-balanced difficulty-weighted replay}\xspace}
\newcommand{\classbalanced}{{class-balanced difficulty-weighted replay}\xspace}

\newcommand{\selectivegeneration}{{selective query}\xspace}
\newcommand{\SelectiveGeneration}{{Selective Query}\xspace}
\newcommand{\Selectivegeneration}{{Selective query}\xspace}

\newcommand{\ours}{{CaBaGE}\xspace}

\newcommand{\clone}{{clone}\xspace}
\newcommand{\Clone}{{Clone}\xspace}
\newcommand{\victim}{{victim}\xspace}
\newcommand{\Victim}{{Victim}\xspace}

\newcommand{\CIFARTVic}{95.54}

\newcommand{\CIFARHslVic}{77.52}

\newcommand{\CIFARTsl}{94.02}
\newcommand{\CIFARTslCI}{0.25}

\newcommand{\CIFARHsl}{69.47}
\newcommand{\CIFARHslCI}{0.88}

\newcommand{\CIFARThl}{87.93}
\newcommand{\CIFARThlCI}{1.74}

\newcommand{\CIFARHhl}{58.72} 
\newcommand{\CIFARHhlCI}{2.42}

\begin{document}
\title{\ours: Data-Free Model Extraction using ClAss BAlanced Generator Ensemble} 
\titlerunning{\ours: DFME using ClAss BAlanced Generator Ensemble}

\maketitle

 \begin{abstract} 

Machine Learning as a Service (MLaaS) is often provided as a pay-per-query, black-box system to clients. Such a black-box approach not only hinders open replication, validation, and interpretation of model results, but also makes it harder for white-hat researchers to identify vulnerabilities in the MLaaS systems. Model extraction is a promising technique to address these challenges by reverse-engineering black-box models. Since training data is typically unavailable for MLaaS models, this paper focuses on the realistic version of it: data-free model extraction.
We propose a data-free model extraction approach, \emph{\ours},  to achieve higher model extraction accuracy with a small number of queries. 
Our innovations include (1) a novel experience replay for focusing on difficult training samples; (2) an ensemble of generators for steadily producing diverse synthetic data; and (3) a selective filtering process for querying the victim model with harder, more balanced samples. 
In addition, we create a more realistic setting, for the first time, where the attacker has no knowledge of the number of classes in the victim training data, and create a solution to learn the number of classes on the fly.  
Our evaluation shows that \ours outperforms existing techniques on seven datasets---MNIST, FMNIST, SVHN, CIFAR-10, CIFAR-100, ImageNet-subset, and Tiny ImageNet---with an accuracy improvement of the extracted models by up to 43.13\%. Furthermore, the number of queries required to extract a clone model matching the final accuracy of prior work is reduced by up to 75.7\%. 

\end{abstract}
 \section{Introduction}


MLaaS~\cite{ribeiro2015mlaas} has seen rapid growth, where a provider offers limited  access, e.g. via Application Programming Interfaces (API), to a machine learning system at a cost. This is known as a pay-per-query system~\cite{jagielski2020high}. Many MLaaS systems are \emph{black-boxes} to the clients. For example, the clients have no knowledge of the model architecture, training method, or the data used to train the model.

Thus, research results built on top of  black-box MLaaS models could be difficult to reproduce,  validate, or interpret, which harms scientific development. 
In addition, it is hard for white hat researchers to identify vulnerabilities and issues in such deployed black-box models. These problems faced by MLaaS clients incentivize the development of model extraction techniques~\cite{jagielski2020high, disguide,dfme,dfms}, i.e., techniques that  steal the MLaaS models. Such models can be used for  reconnaissance to launch further attacks~\cite{papernot2017practical,Shokri_2017_Membership_Inference_Attacks}.


Existing research focuses on learning-based model extraction~\cite{dfme,disguide,dfms,zhang2023ideal},
the process of using only information gained by querying a black-box model, the \textit{victim}, to train a machine learning system, the \textit{clone}, for application on the same task. However, the victim training data is often inaccessible, and constructing a surrogate dataset for training is a difficult and expensive task \cite{dfme}. This prompts researchers to borrow ideas from Generative Adversarial Networks (GAN), and use generative models to generate synthetic data for querying the victim model. In this paper, we follow this generative approach and assume the following constraints to make the extraction process ``data-free''. First, the attacker knows only the {\victim}'s input data format and has no further information pertaining to the training data or the target system. The second assumption is that the attacker has no access to any data that can be used in a comparable format for evaluation or training purposes. 

Since data-free model extraction is a difficult problem, much of the prior work has focused on more accessible variants of the problem. These often involve either high ($\geq 8$ million) query budgets or the assumption that the target model offers exact model confidence values, known as \textit{soft-label} (SL) extraction~\cite{disguide,dfme,dfms,Kariyappa_2021_CVPR}. Yet, in a pay-per-query system, a high query budget means the technique is prohibitively expensive or impractical. Additionally, any attack relying on exact model confidence values can be countered by the trivial defence of giving only the top-1 or top-k label predictions. To address the countermeasure of providing only the top-k label predictions, previous studies~\cite{dfms,disguide,zhang2023ideal} have adopted \textit{hard-label} (HL) extraction, where the victim model returns only the top-1 label prediction.
Prior papers make the assumption that the total number of classes is known.



To make the extraction as realistic as possible, our extractor for HL extraction is assumed to have no knowledge of the number of classes, and must learn the number of classes in the target domain. This \emph{class-agnostic} setting is a stricter, i.e., more realistic, assumption than the class-aware setting used by prior work. To the best of our knowledge, this is the first work where the attacker knows only the input data format (i.e., images and their dimensions) and the general goal of the task, in this case, image classification.

We propose a novel, data-free model extraction approach---\emph{\ours}---that combines three key techniques: \classbalanced, generator ensemble, and \selectivegeneration. \Classbalanced balances the class distribution of the replayed samples and leverages priority sampling to keep the most difficult samples when the memory is full. Generator ensemble uses an ensemble of generators with increased generator training iterations to generate more diverse data in data-free model extraction. \Selectivegeneration is a filtering process to select balanced samples to query the victim. 

This paper answers three key questions:


\begin{enumerate}
    \item Does \ours achieve a higher accuracy than existing approaches?
    \item Does \ours achieve a higher query efficiency than existing approaches?
    \item How do each of \ours's novel components contribute to the results?
\end{enumerate}


\newpage
\noindent
Our work makes the following contributions:
\begin{itemize}
    \item We propose a novel data-free model extraction approach \emph{\ours} that combines three key techniques: \classbalanced, generator ensemble, and \selectivegeneration. \ours generates and selects more diverse, balanced, and higher-quality data for data-free model extraction to achieve higher extracted accuracy with fewer number of queries. 
    \item We create a realistic class-agnostic setting, for the first time,  where the attacker has no knowledge of the number of classes in the victim training data, and must instead learn the number of classes on the fly. \ours adaptively modifies the prediction head of the clone models based on the labels obtained from querying the victim model. For all HL evaluations in this paper, \ours uses the more challenging, realistic class-agnostic setting, while existing work uses the class-aware setting. 
    \item Our evaluation shows that in limited-budget settings, \ours outperforms the State-of-The-Art (SoTA) techniques, DisGUIDE and IDEAL, on all seven datasets---MNIST, FMNIST, SVHN, CIFAR-10, CIFAR-100, ImageNet-subset, and Tiny ImageNet. On simpler datasets such as MNIST, FMNIST, and SVHN, \ours improves the final accuracy by up to $43.13\%$, $37.09\%$, and $9.04\%$ respectively. On the more complex datasets, CIFAR-100 and ImageNet subset, \ours  achieves $11.10\%$ and $26.23\%$ gains in final accuracy.
   
    \item For a fair comparison with DisGUIDE, we also evaluate \ours following DisGUIDE's extraction configuration, which assumes a higher query budget. In this setting, \ours's HL extraction performance outperforms the best final accuracy of DisGUIDE on CIFAR-10 and CIFAR-100 by $1.35\%$ and $5.73\%$. In the SL setting, we observe similar gains, improving final accuracy by $0.34\%$ and $6.49\%$ on CIFAR-10 and CIFAR-100 models respectively. Most significantly, \ours achieves a leap in accuracy to reach 75.96\% out of a 77.52\% victim, on CIFAR-100 in the SL setting.
\end{itemize}
 \section{Related Work}
\label{relatedwork}
\subsection{Model Extraction}
In the context of machine learning, model extraction is a class of attacks whereby an adversary with black-box access to a machine learning system seeks to obtain valuable information of the model, which includes: training hyperparameters, learned parameters, or an approximation of the model with a high agreement over relevant input spaces~\cite{197128,oliynyk2022know}.\\
\textbf{Data-Free Model Extraction}:
In a more challenging scenario of model extraction, the authors of DFME~\cite{dfme} assume that adversaries have no access to initial training data. Instead, DFME trains a substitute (clone) model on synthetic data generated by a GAN-like mechanism, and queries the target model for class predictions to serve as proxy labels. Since the {\victim}s are black-box models, DFME employs a forward differences method to approximate the necessary gradients for clone training. However, this approach requires many queries to estimate gradients, and suffers in HL extractions. Subsequent studies including DFMS~\cite{dfms}, DisGUIDE~\cite{disguide}, and IDEAL~\cite{zhang2023ideal} have expanded on DFME's work. DFMS proposes training a GAN to emulate synthetic or real data while maximizing clone label confidence entropy. DisGUIDE introduces the use of replay methods and utilizes a generator training loss that calculates the difference in clone models' prediction to make queries more efficient. IDEAL pushes model extraction further towards low query budget settings by querying generated samples with the highest clone confidence.

\subsection{Model Distillation}
Model distillation, also known as knowledge distillation, refers to the transfer the knowledge from a teacher model to a smaller student model~\cite{hinton2015distilling, gou2021knowledge}. 
Different from model extraction, there is typically white-box access to the teacher model and the emphasis is often on the performance relative to parameter size of the resulting trained model or the amount of arithmetic operations required in the training process. This differs from model extraction, where the attacker only has black-box access to the victim and aims to reduce the number of victim queries to reach high fidelity or accuracy in the targeted domain~\cite{jagielski2020high}.
\\
\textbf{Data-Free Model Distillation}: In the assumption that the teacher's training data is not accessible, i.e., data-free, some existing works aim to capture the distribution of teacher training data by using information stored in teacher model's layers~\cite{lopes2017data,yin2020dreaming}. Other newer approaches borrow ideas from GANs---using a generator to produce training data where the generator's goal is to maximize the disagreement between the teacher and student~\cite{fang2019datafree,Micaelli2019ZeroShotKT}. These approaches attempt to explore and map out the decision boundaries of the teacher to more effectively train the student by querying on the decision boundary. Data-Free Model Extraction borrows these newer ideas, using synthetically generated data to transfer the knowledge of the victims to clones.
\subsection{Ensemble Learning}
\label{sec:2.3}
Ensemble methods can improve the generalization of neural networks and reduce the high variance properties of the models~\cite{GANAIE2022105151}. The general simplicity in implementation and overall improvement of Ensemble Learning methods has resulted in their use across many different machine learning and deep learning fields~\cite{Yuksel2012_Mixture_of_Experts}. \\
\textbf{{Ensemble Learning in Adversarial-Learning}}: Prior work shows that training a GAN with an ensemble of generators improves performance and the diversity of the generated outputs~\cite{hoang2018mgan,Ghosh2018madGAN}. Existing work in data-free model extraction also utilizes the concept of ensemble learning. Rosenthal et al. leverage an ensemble of clones to improve the stability of clones prediction~\cite{disguide}. Others deploy two generators as an ensemble and optimize the disagreement of the generator outputs, trying to boost the diversity of the generated outputs~\cite{related-generator-ensemble-DFME}. In contrast, \ours{} does not directly compare generator outputs, but relies on the joint optimization process to incentivize ensemble diversity implicitly. In addition, our approach has a negligible increase in computational cost (details in~\cref{sec:ge}).

 \section{Approach}
\label{methods}

\begin{figure}[t] 
    \centering
    \includegraphics[width=0.8\textwidth]{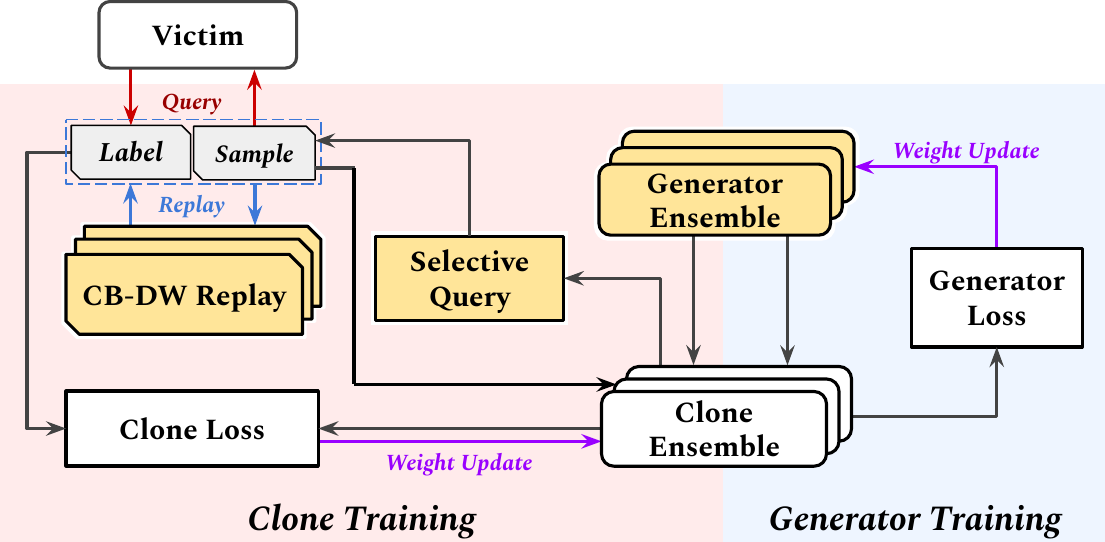} 
    \caption{Overview of \ours. Our three novel components are colored in yellow.\\ CB-DW Replay is the \ClassBalanced.}
    \label{fig:overview} 
\end{figure}
\cref{fig:overview} is an overview of \ours. \ours' three novel components are highlighted in yellow: Generator Ensemble (\cref{sec:ge}), \SelectiveGeneration (\cref{sec:sq}), and \ClassBalanced (\cref{sec:replay}).
We build \ours upon the foundational method introduced by DisGUIDE~\cite{disguide}. In DisGUIDE, an attacker trains a generator and two clones in an adversarial-like setting and the final extracted model is derived from the ensemble of the clones.

Similar to DisGUIDE, \ours's extraction process is composed of two phases, shown in~\cref{fig:overview}: (1) \textit{\Clone Training}, and (2) \textit{Generator Training}. Within the given query budget, \ours cycles between these phases. Before entering any phases, \ours's extraction process starts by initializing the \textit{generator ensemble} and the \textit{clone ensemble} from random weights. Afterwards, \ours optimizes the generator ensemble's weights in the \textit{Generator Training} phase, while keeping the \textit{clone ensemble}'s parameters frozen. 
Conversely, the \textit{\Clone Training} phase updates the \clone ensemble while keeping the generator ensemble fixed.

The \textit{\Clone Training} phase has two stages. First, we apply \SelectiveGeneration, our proposed filtering technique, to generate samples and select a final batch to query the victim with. The clones then use these samples and the returned labels for training. The samples and labels are then stored in the \ClassBalanced. In the second stage of \textit{\Clone Training}, we sample from the \ClassBalanced and train the clones without querying the victim. We introduce three key improvements in \ours:

\begin{enumerate}
    \item \textit{Generator Ensemble}: We propose to train an ensemble of generators instead of a single one. Ensembling the generator prevents mistakes in a single generator from poisoning the sole source of data for the extraction. Compared to the single generator approach used in DisGUIDE, our proposed solution requires neither additional queries to the \victim nor extra computation.
    \item \textit{\SelectiveGeneration}: During \textit{\Clone Training}, we propose a strategy for choosing harder, class-balanced samples to query the \victim with. This helps to improve the stability of the extraction without increasing the query budget.
    \item \textit{\ClassBalanced}: Prior work generally fails to efficiently utilize the replay method, which is crucial in the DFME context where our only data sources are those we have queried. We thus propose an improvement to the replay memory of prior work~\cite{disguide}, by both balancing the class distribution of returned samples and keeping more difficult samples for the clones as memory fills up. 
\end{enumerate}

\subsection{Generator Ensemble}
\label{sec:ge}

To prompt diversity in the generated data and a more stable extraction performance in the DFME-HL setting, we propose the use of generator ensembles. Traditionally, only a single generator \( G \) is used to capture the entire data distribution \( p_{\text{data}} \). Instead, we propose the use of $G_E$, an ensemble of \( n \) generators \( \{G_1, G_2, \dots, G_n\} \), to capture the data distribution. In this approach, for a batch of data samples \( B \) to be generated, each sample  within the batch will be associated with an index and is denoted by \( x_i \). The batch is partitioned into \( n \) sub-batches such that each generator \( G_j \) in the \textit{generator ensemble} $G_E$ is responsible for generating samples corresponding to the indices in $S_j$, its assigned sub-batch. The batch size remains fixed irrespective of generator ensemble count. Formally, if \( S_j \) represents the set of indices assigned to \( G_j \), then $x_i$, the image output of \( G_j \) for a corresponding input noise vector \( z_i \) is: 
    \begin{equation}
    \small
    x_i \leftarrow G_j(z) \text{ for } i \in S_j 
    \end{equation}
Each time \ours enters the generator training phase, the whole \textit{generator ensemble} is trained by one batch. Within this batch, each generator model \( G_j \) is trained to maximize both the diversity and clone ensemble disagreement on its own generated image data, $\{x_i\}_{i\in S_j}$.

We note that the diversity of the generated data is a global objective shared between ensemble members. If one generator creates too many instances of a given class, other ensemble members can compensate by creating fewer instances of that class. On the other hand, the clone ensemble disagreement on a given sample is independent of other samples in a given batch.

During \textit{\Clone Training}, all generator outputs are combined to produce a full image batch of size $|B|$, where $|B| \, / \, n$ images are produced by each generator in the ensemble. Due to this design, our ensembling approach incurs no extra computational cost in the forward and backward propagation.

This approach benefits from two advantages:

    \noindent\textit{Specialization}: Each generator specializes in a subset of the data distribution, reducing the complexity of what individual generators must learn. 
    
    \noindent\textit{Diversity}: Multiple generators may enhance the ability to cover the full targeted data distribution and help to mitigate the risk of mode collapse, a common problem faced by GAN-like methods.

\subsection{\SelectiveGeneration}
\label{sec:sq}

To query the \victim with more challenging images in \textit{\Clone Training}, we introduce a selection process for newly generated data, \textit{\SelectiveGeneration}. This method is based on the belief that querying the victim with diverse and difficult samples benefits clone training. \textit{\SelectiveGeneration} is a three-step process:
\begin{enumerate}
    \item \textit{Oversampling}: Generators produce multiple batches of images.
    \item \textit{Evaluation}: The clone ensemble evaluates every sample in the over-sampled data pool. Two primary metrics are computed for each input based on these outputs: The ensemble disagreement loss and the class label.
    \item \textit{Selection}: Select an equal number of images from each class, determined by its predicted class label using the averaged class probabilities of the clones, prioritizing those with the highest disagreement loss. This ensures the final selected data is balanced and challenging.
\end{enumerate}

Problems arise when limited or no data generated is predicted to be of certain classes. To address this situation, missing samples are replaced in two steps. First, half the missing samples are chosen by selecting from the remaining images with the highest disagreement loss, regardless of their classes. The other half of the missing samples are drawn uniformly from the residual image samples. \textit{\SelectiveGeneration} aligns with the intuition of~\cite{10.1109/ICSE48619.2023.00111} and~\cite{hu-etal-2016-harnessing}, in which the generated data is constrained with specific rules at both training and inference time. A detailed algorithm is provided in supplementary materials.

\subsection{\ClassBalanced}
\label{sec:replay}

 We draw inspiration from work that is shown to be promising in the Continual Learning domain~\cite{9412614}, and create a simple yet effective memory replay. This replay has the following primary features.

 \begin{enumerate}
     \item The replay yields class-balanced samples.
     \item When the allocated space for a particular class is saturated, replacement skews towards easier samples: those with a lower \clone training loss.
 \end{enumerate}

For each respective class \(k\) the attacker discovers, we initialize a separate class memory bank, denoted as \(M_k\). Samples are stored to \(M_k\) if and only if the \victim classifies it as belonging to that class. The maximum capacity of each memory bank is equal to a fixed total memory size divided by the number of classes the attacker has discovered. The class memory banks are managed by a container \(M\) which ensures samples are stored correctly and sampled evenly. During training from replay, an equal number of samples from each respective class known to the attacker are sampled from the class memory banks. This simple improvement reduces class imbalance.

As the memory bank fills up, samples eventually need to be removed to make space for new ones. We follow inspiration from~\cite{9412614}: using a weighted random sampling to select samples for replacement. This aims to keep the most valuable data in storage. We compute the weighting based on the most recent inverse \clone training loss value for each respective sample. For each batch of samples we update, we apply a transformation where we subtract loss from the maximum loss value within the batch. In this way, samples that are harder for the clones to learn are more likely to be retained for longer in the memory bank.

To the best of our knowledge, previous works on data-free model extraction that employ a memory bank for experience replay rely on simpler methods, such as a circular buffer for memory storage or random sampling strategies~\cite{disguide, Binici_2022_WACV}, which do not consider the importance of class balance or the difficulty of samples.

\subsection{Loss Functions}
The equation below describes the \clone training loss $L_C$ for clone $c_i$ in the HL setting. $K$ represents the classes we have discovered up to this point of {\ours}'s extraction process, and $c_i\left(s_n\right)_{k}$ represents the logit of clone $c_i$ on class $k$ for generated sample $s_n$. Finally, $p_V$ is label assigned by querying the \victim.

\begin{equation}
\small
HL:\ \ L_C = -\frac{1}{N} \sum_{n=1}^{N} \log \left( \frac{{\exp(c_i\left(s_n\right)_{p_V})}}{{\sum_{k\in K} \exp(c_i\left(s_n\right)_{k})}} \right)
\end{equation}

For SL extraction, the \clone training loss is the MSE loss computed between the pseudo logits of the victim and the clones' raw logits. Here, we follow DFME's approach to obtain the pseudo logits $V(s_n)$ of the victim~\cite{dfme}.
\begin{equation}
\small
SL:\ \ L_C = -\frac{1}{N} \sum_{n=1}^{N} \left( c_i(s_n)-V(s_n) \right)^2
\end{equation}

{\ours}'s generator training's optimization goal follows previous work, in which the generator jointly optimizes the \textit{disagreement loss} $L_D$, which aims to maximize the disagreement between clone models, and the \textit{class diversity loss} $L_{div}$, aiming for a balanced data distribution in the generated data. $\lambda$ is a hyper-parameter used as a weighting coefficient for the {class diversity loss}.

\begin{equation}
\small
L_G = L_D + \lambda L_{div}
\end{equation}

Following previous work~\cite{disguide} and~\cite{addepalli2020degan}, the disagreement loss $L_D$ is the standard deviation of the fixed clone models' prediction over all previously discovered classes, and $L_{{div}}$ is the information entropy of the clones' prediction.  

 \section{Experimental Setup}
\label{experimental-setup}

To compare with prior work in data-free model extraction, we follow their experimental configurations. Specifically, we examine two settings, the first from IDEAL~\cite{zhang2023ideal}, and the other from DisGUIDE~\cite{disguide}. We denote the settings from IDEAL, where query budgets are much lower, as the \textit{limited-budget setting} ($\leq 2M$ queries ), and the setting from DisGUIDE as the \textit{relaxed-budget setting} ($\geq 8M$ queries). We perform extraction from the following seven datasets: MNIST~\cite{lecun1998gradient}, FMNIST~\cite{xiao2017fashionmnist}, SVHN~\cite{netzer2011reading}, CIFAR-10~\cite{krizhevsky2009learning}, CIFAR-100~\cite{krizhevsky2009learning}, Tiny ImageNet~\cite{le2015tiny} and an ImageNet-subset~\cite{li2021anti}. For each dataset, we extract 2 or 3 victim architectures, dependent on the dataset. The list of model architectures is: MLP, LeNet~\cite{lecun1998gradient}, AlexNet~\cite{krizhevsky2012imagenet}, VGG-16~\cite{Simonyan15}, ResNet-18~\cite{he2016deep}, and ResNet-34~\cite{he2016deep}. We use IDEAL's published implementations of MLP, LeNet, AlexNet, ResNet-18, and ResNet-34 architectures~\cite{zhang2023ideal}, and DFME's published implementation for VGG-16~\cite{dfme}. 
To eliminate any potential biases, we run all extraction techniques on the same victim models for each setting in which we make comparisons. The victims and their training details are specified in the supplementary materials.

\noindent \textbf{Evaluation Metric:} To compare with prior work in DFME~\cite{zhang2023ideal,disguide}, we focus our evaluation on accuracy. Since researchers are also interested in fidelity results~\cite{jagielski2020high}, we refer readers to the supplementals for \ours{'s} fidelity results.

\subsection{Limited-Budget and Relaxed-Budget Setting}
\label{exp_setting_limited}
In the limited-budget setting, we compare three techniques: IDEAL, DisGUIDE, and \ours. Following IDEAL's settings, we use 25K queries for extractions on MNIST, 100K for FMNIST and SVHN, 250K for CIFAR-10 and ImageNet-subset, and 2M for CIFAR-100 and Tiny ImageNet. ~\cref{tab:qblimited_accuracy} reports the average final accuracies obtained by each respective method. On CIFAR-100 and Tiny ImageNet the clone model architecture is ResNet18, while in all other cases the clone architecture is congruent to the victim model architecture.

For reproducibility, we use the publicly accessible repositories of IDEAL\footnote{\url{https://github.com/SonyResearch/IDEAL/tree/main}} and DisGUIDE\footnote{\url{https://github.com/lin-tan/disguide}}. However, there is a discrepancy between the query budget definitions in the IDEAL paper~\cite{zhang2023ideal} and its code base. 
IDEAL's code base queries the victim model with multiple different images, that are augmented versions of each other, but the IDEAL paper only counts these multiple queries as a single query. Due to this discrepancy, IDEAL's extraction experiments query the victim model multiple times more than the reported query budget in their paper. These additional queries represent a budget increased by a factor of 2 times for non-CIFAR or 162 times for CIFAR datasets. To ensure an equitable comparison in the limited-budget setting, we report reproduced results of IDEAL without mutating the stored images to make the query budget constraint consistent with what is described in the IDEAL paper. We denote this query budget adjusted version of IDEAL as IDEAL$^*$. The details of this discrepancy and IDEAL$^*$ are provided in the supplementary materials.

In the relaxed-budget setting, we compare \ours with DisGUIDE in both SL and HL extraction on CIFAR-10 and CIFAR-100 datasets. The experiments follow DisGUIDE's exact experimental settings, details can be found in~\cite{disguide}.

\subsection{Hyper-Parameters}
In a realistic setting for DFME, the attacker is only aware of the model architecture they have selected to replicate the target system, as well as the query budget. Thus, most hyper-parameters should be identical across all experiments. Consequently, we choose to fix the learning rate for clone models with the same architecture and with the same query budget. We set the learning rate for AlexNet clones to be fixed at 0.004, VGG-16 clones at 0.01, ResNet18 clones at 0.03, and ResNet34 clones at 0.1, respectively, for all query budgets. For MLP clones, learning rates are 0.01 (25K queries) and 0.0125 (100K queries). For LeNet clones, they are 0.1 (25K queries) and 0.01 (100K queries). 
 
For all experiments, \ours uses a batch size of 250. We use a clone ensemble size of $2$, and a generator ensemble size of $8$.
Within each Generator Training phase, we train the generator for 3 batches. In the Clone Training phase, Selective Query selects a batch from $1000$ samples to query the victim with, and the clone models are trained for 1 iteration using the newly queried batch. The clone model is trained with 12 batches of data sampled from the knowledge replay. Learning rates are scheduled to drop by 0.3 at 40\% and 80\% of the total query budget under the relaxed-budget setting. Learning rate drops were not used in the limited-query budget setting. Additionally, as the diversity loss weighting should increase based on the number of discovered classes~\cite{disguide}, we dynamically scale $\lambda$ during training, varying it approximately inversely with the number of discovered classes via a simple relation: $\lambda = \frac{4}{(10 + K)}$. For all experiments, we follow DisGUIDE~\cite{disguide}, and use a replay size of 1 Million.

\begin{table}[t]
    \centering
    \scriptsize
    \caption{Accuracy (\%) and 95\% confidence interval of \clone in various limited-budget settings. Experiments result for DisGUIDE and \ours are computed over 3 runs. IDEAL$^*$'s result are run by a fixed random seed based on the published code. 
    IDEAL$^*$ is the query budget adjusted version of IDEAL for a fair comparison (~\cref{exp_setting_limited}).
    }
    \begin{tabular}
    {@{\hspace{3pt}}l@{\hspace{10pt}}l@{\hspace{6pt}}c@{\hspace{5pt}}c@{\hspace{10pt}}c@{\hspace{15pt}}c@{\hspace{7pt}}c}
    \toprule
    \textbf{Dataset} & \textbf{Model} & \textbf{Victim} & \textbf{IDEAL$^*$} & \textbf{DisGUIDE} & \textbf{\ours} \\
    \midrule
      & MLP & 98.25 & 14.00 & $11.35\pm0.00$ & $\boldsymbol{57.13}\pm4.15$ \\
MNIST & LeNet & 99.27 & 86.40 & $91.23\pm2.01$ & $\boldsymbol{94.49}\pm0.81$ \\
      & AlexNet & 99.35 & 66.30 & $18.11\pm5.23$ & $\boldsymbol{85.31}\pm0.85$ \\
\midrule
       & MLP & 84.54 & 19.00 & $37.53\pm5.75$ & $\boldsymbol{74.62}\pm3.15$ \\
FMNIST & LeNet & 90.23 & 27.80 & $59.43\pm2.86$ & $\boldsymbol{68.72}\pm2.39$ \\
       & AlexNet & 92.66 & 35.20 & $54.47\pm5.29$ & $\boldsymbol{81.45}\pm0.64$ \\
\midrule
     & VGG-16 & 94.41 & 68.35 & $79.96\pm3.21$ & $\boldsymbol{84.10}\pm0.36$ \\
SVHN & ResNet-18 & 95.28 & 72.60 & $75.83\pm0.77$ & $\boldsymbol{78.09}\pm0.84$ \\
     & AlexNet & 89.82 & 67.00 & $19.39\pm6.70$ & $\boldsymbol{76.04}\pm2.42$ \\
\midrule
\multirow{2}{*}{CIFAR-10} & AlexNet & 84.76 & 25.50 & $24.73\pm3.23$ & $\boldsymbol{33.35}\pm1.63$ \\
         & ResNet-34 & 93.85 & 20.40 & $18.05\pm5.97$ & $\boldsymbol{26.03}\pm2.58$ \\
\midrule
\multirow{2}{*}{CIFAR-100} & AlexNet & 63.38 & 6.17 & $24.45\pm0.38$ & $\boldsymbol{33.09}\pm0.66$ \\
          & ResNet-34 & 77.52 & 7.94 & $32.65\pm0.64$ & $\boldsymbol{43.75}\pm1.75$ \\
\midrule
\multirow{2}{*}{ImageNet-subset} & AlexNet & 72.96 & 20.60 & $18.92\pm0.91$ & $\boldsymbol{46.83}\pm3.06$ \\
                & VGG-16 & 78.53 & 20.50 & $31.01\pm2.09$ & $\boldsymbol{37.04}\pm7.60$ \\
\midrule
\multirow{2}{*}{Tiny Imagenet} & ResNet-34 & 59.28 & 4.93 & $11.87\pm2.73$ & $\boldsymbol{15.36}\pm0.64$ \\
             & VGG-16 & 42.04 & 4.15 & $7.87\pm1.50$ & $\boldsymbol{11.98}\pm0.64$ \\

    \bottomrule
    \end{tabular}
    \label{tab:qblimited_accuracy}
\end{table}

\section{Results}
\subsection{Extraction Accuracy}
\label{sec:5.1}
We compare the performance of our method, \ours,  with two SOTA data-free model extraction techniques: DisGUIDE~\cite{disguide} and IDEAL~\cite{zhang2023ideal}. IDEAL emphasizes extraction under stringent query budget constraints, while DisGUIDE aims at high-performance extraction with a more generous budget. We report the accuracy in extraction settings matching both of these prior papers respectively.

\vspace{5pt}
\textbf{Limited-Budget Setting}
\cref{tab:qblimited_accuracy} shows the extraction results of IDEAL$^*$, DisGUIDE, and \ours in terms of final extracted accuracy and 95\% confidence interval under the limited-budget setting, described in~\cref{exp_setting_limited}. 
IDEAL$^*$ is the query budget adjusted version of IDEAL, for a fair comparison as described in~\cref{exp_setting_limited}. For example, when extracting the MLP victim trained on MNIST, \ours reaches $57.13\%$ accuracy on the test set, outperforms the prior best extraction result of 14.00\% by 43.13\%. Consistently, \ours outperforms prior work on all victims, on any target dataset.

For elementary victims trained on less intricate datasets, such as MNIST and FMNIST, \ours demonstrates significant performance improvements, improving the final test-set accuracy by an average of $23.00\%$ for AlexNet, $6.27\%$ for LeNet, and $40.11\%$ for MLP. More specifically, \ours extracts a LeNet model trained on MNIST with an averaged final accuracy of 94.49\%, which is very close to the victim model's accuracy of 99.27\%.
On slightly more complex datasets including SVHN and CIFAR-10, \ours also exhibits improvements over previous SOTA. Extraction on SVHN datasets achieves on average a $5.15\%$ increase, reaching $79.41\%$, and extraction on CIFAR-10 achieves on average a $6.74\%$ increase, reaching $29.69\%$.
Similar improvements are also shown in extracting victim models trained on more challenging datasets. Namely, on the ImageNet-subset, CIFAR-100, and Tiny ImageNet datasets, we observe mean improvements of $16.13\%$, $9.87\%$, and $3.80\%$ respectively. 

\smallskip\noindent\fbox{
\centering
\parbox{0.95\linewidth}{
\textbf{Summary}: In the limited-budget setting, \ours outperforms both DisGUIDE and IDEAL$^*$ across all 17 settings, increasing accuracy up to $43.13\%$.
}}

\subsubsection{Relaxed-Budget Setting}

\begin{table}[t]
\centering
\scriptsize
\caption{Final clone accuracy comparison with 95\% confidence intervals in the relaxed-budget setting.
DisGUIDE~\cite{disguide} 
results are the paper reported accuracies. 
}
\begin{tabular}{ll rr@{ }l rr@{ }l}
 \toprule
 \multirow{2}{*}{\textbf{Setting}} & \multirow{2}{*}{\textbf{Technique}} 
 & \multicolumn{3}{c}{\textbf{CIFAR-10}} & \multicolumn{3}{c}{\textbf{CIFAR-100}} \\
 
 \cmidrule(lr){3-5} \cmidrule(lr){6-8}
 
 &
 & \textbf{Victim}  & \multicolumn{2}{r}{\textbf{Clone (\%)}} 
 & \textbf{Victim}  & \multicolumn{2}{r}{\textbf{Clone (\%)}}\\
 
 \midrule
 \multirow{2}{*}{Soft Label \hspace{5pt}} 
 & DisGUIDE  & \CIFARTVic & \CIFARTsl & $\pm$ \CIFARTslCI & \CIFARHslVic & \CIFARHsl & $\pm$ \CIFARHslCI \\ 
 & \textbf{\ours}  & 95.54 & \textbf{94.36} & $\pm$ 0.05   & 77.52 & \textbf{75.96} & $\pm$ 0.25 \\ 
 \midrule
 \multirow{2}{*}{Hard Label} 
 & DisGUIDE  & \CIFARTVic &  \CIFARThl & $\pm$ \CIFARThlCI  & \CIFARHslVic & \CIFARHhl & $\pm$ \CIFARHhlCI \\
 & \textbf{\ours}  & 95.54 & \textbf{89.28} & $\pm$ 0.61 & 77.52 & \textbf{64.45} & $\pm$ 0.87 \\

 \bottomrule
\end{tabular}
\label{tab:method_comparison_acc}
\vspace{-2mm}
\end{table}

\label{sufficient_qb_acc_comparison}
\cref{tab:method_comparison_acc} compares the final accuracy of \ours on CIFAR-10 and CIFAR-100 following DisGUIDE configurations in the relaxed-budget setting for both SL and HL extractions. For SL extractions, when extracting from the ResNet-34 victim trained on CIFAR-100, \ours outperforming DisGUIDE by $6.49\%$, achieving a final accuracy of $75.96\%$ on the test set, which is $97.99$ percent of the victim model's accuracy ($77.52\%$). This underscores that with a higher query budget, \ours offers highly accurate SL model extraction, even on more intricate models trained with complex datasets. \ours also increases the final extracted accuracy on CIFAR-10 by $0.34\%$. Similarly, in the HL setting, the test-set accuracies are increased by  $1.35\%$ and $5.73\%$ for extraction processes on CIFAR-10 and CIFAR-100, reaching $89.28\%$ and $64.45\%$ accuracy respectively. Besides the gains of final accuracy of the extracted models, \ours also makes the extraction process more stable, i.e., reduces the fluctuation of results, when the query budget is relaxed. 

\smallskip\noindent\fbox{
\centering
\parbox{0.95\linewidth}{
\textbf{Summary}: Under the relaxed-budget setting, \ours outperforms prior work on CIFAR-10 and CIFAR-100. In the HL setting, \ours improves the final test-set accuracy by $1.35\%$ and $5.73\%$. Our biggest gain is in the SL setting, where \ours achieves an accuracy improvement of $6.49\%$, reaching $97.99\%$ of the victim model's accuracy on CIFAR-100.}}\smallskip

\subsection{\ours's Query Efficiency}

\begin{table}[t]
\scriptsize
\centering
\caption{Mean number of queries (in millions) to reach prior work---DisGUIDE's reported final accuracies with respective 95\% confidence intervals. Lower is better.}
\begin{tabular}{l cc cc}

\toprule
\multirow{2}{*}{\textbf{Setting}} 
& \multicolumn{2}{c}{\textbf{CIFAR-10}} & \multicolumn{2}{c}{\textbf{CIFAR-100}}\\

\cmidrule(lr){2-3} \cmidrule(lr){4-5}

& \hspace{2pt} \textbf{DisGUIDE} & \textbf{\ours} 
& \hspace{2pt} \textbf{DisGUIDE} & \textbf{\ours}\\
\midrule
 {Soft Label}
 & \hspace{2pt} 20M & \textbf{16.04M} $\pm$ 0.01M & \hspace{2pt} 10M & \textbf{2.43M} $\pm$ 0.15M  \\
 \midrule
 Hard Label
 & \hspace{4pt} 8M  & \hspace{2pt} \textbf{6.32M} $\pm$ 0.13M & \hspace{2pt} 10M & \textbf{6.07M} $\pm$ 1.81M \\
 \bottomrule
\end{tabular}
\label{tab:query_efficiency}
\vspace{-5mm}
\end{table}

We compare the number of queries required for \ours to reach the final accuracy of the prior work DisGUIDE against its query budgets, and report the result in~\cref{tab:query_efficiency}. In all extraction settings, \ours shows better query efficiency compared to DisGUIDE. For SL extraction on CIFAR-100\, \ours needs only 2.43 million queries on average to achieve similar accuracy to DisGUIDE's final result, with a 95\% confidence interval of 0.15 million, reducing the prior standard by 75.7\%. For SL extraction on CIFAR-10 the number of queries needed to reach prior work's best is reduced by 3.96 million on average, 19.8\% of the total queries.

In the HL settings, \ours reaches the prior SOTA accuracy in 6.32 million queries on CIFAR-10 and with 6.07 million queries on CIFAR-100. This constitutes a reduction in the needed queries by 21\% and 39.3\%, respectively.

\smallskip\noindent\fbox{
\centering
\parbox{0.95\linewidth}{
\textbf{Summary}: \ours consistently improves query efficiency in both the CIFAR-10 and CIFAR-100 datasets under SL and HL settings. Extractions on the CIFAR-100 dataset, in particular, show a reduction in the number of required queries in the SL setting by \textbf{75.7\%}.
}}\smallskip

\subsection{Ablation  Study} 

\begin{table}
\centering
\scriptsize
\caption{Final Accuracy of \ours and ablation settings shown with 95\% confidence interval. Independent t-tests are used to determine improvement over the baseline DisGUIDE$^*$, where statistically significant improvements are highlighted in \color{blue}{Blue}\color{black}.
}

\begin{tabular}{l@{\hspace{14pt}}l@{\hspace{14pt}}rl@{\hspace{14pt}}rl@{\hspace{14pt}} 
rl@{\hspace{14pt}}rl}

\toprule

\multirow{2}{*}{\textbf{Dataset}} & \multirow{2}{*}{\textbf{Model}} & \multicolumn{2}{l}{\hspace{-3pt}\textbf{DisGUIDE$^*$}} & \multicolumn{2}{l}{\hspace{-3pt}\textbf{DisGUIDE$^*$}} & \multicolumn{2}{l}{\hspace{-3pt}\textbf{DisGUIDE$^*$}} & \multicolumn{2}{c}{\multirow{2}{*}{\textbf{\ours}}} \\

 & & \multicolumn{2}{c}{} & \multicolumn{2}{l}{\hspace{1pt}\textbf{$_\text{+CB-DW}$}} & \multicolumn{2}{l}{\hspace{-6pt}\textbf{$_\text{+CB-DW+GE}$}} & \multicolumn{2}{c}{} \\
    
    \midrule
\multirow{3}{*}{MNIST} & MLP & $11.35$&$\pm0.00$ & \color{blue}{$48.58$}&$\pm5.73$ & \color{blue}{$54.32$}&$\pm4.04$ & \color{blue}{$54.94$}&$\pm3.58$ \\
& LeNet & $94.20$&$\pm0.55$ & $94.02$&$\pm0.88$ & $94.84$&$\pm1.07$ & \color{blue}{$95.32$}&$\pm0.63$ \\
& AlexNet & $74.10$&$\pm2.11$ & $71.64$&$\pm4.31$ & \color{blue}{$82.54$}&$\pm3.30$ & \color{blue}{$86.08$}&$\pm0.94$ \\

\midrule
\multirow{3}{*}{FMNIST} & MLP & $42.28$&$\pm4.40$ & \color{blue}{$72.80$}&$\pm2.85$ & \color{blue}{$76.61$}&$\pm2.35$ & \color{blue}{$74.45$}&$\pm3.28$ \\
& LeNet & $69.12$&$\pm1.12$ & $67.10$&$\pm1.83$ & $70.29$&$\pm1.58$ & $68.08$&$\pm1.50$ \\
& AlexNet & $74.66$&$\pm2.09$ & $75.76$&$\pm2.19$ & \color{blue}{$78.86$}&$\pm1.32$ & \color{blue}{$79.63$}&$\pm1.48$ \\

\midrule
\multirow{3}{*}{SVHN} & VGG-16 & $88.37$&$\pm0.87$ & $86.67$&$\pm2.12$ & 
{$83.63$}&$\pm1.32$ & 
{$84.14$}&$\pm2.09$ \\
& ResNet-18 & $76.05$&$\pm1.81$ & $77.87$&$\pm1.52$ & \color{blue}{$79.49$}&$\pm1.18$ & \color{blue}{$78.70$}&$\pm0.98$ \\
& AlexNet & $64.01$&$\pm2.40$ & \color{blue}{$72.27$}&$\pm2.89$ & \color{blue}{$72.35$}&$\pm4.06$ & \color{blue}{$74.46$}&$\pm2.19$ \\

\midrule
\multirow{2}{*}{CIFAR-10} & AlexNet & $26.25$&$\pm3.45$ & $30.09$&$\pm1.82$ & \color{blue}{$35.66$}&$\pm1.44$ & \color{blue}{$33.23$}&$\pm1.04$ \\
& ResNet-34 & $22.04$&$\pm2.64$ & \color{blue}{$28.78$}&$\pm4.10$ & \color{blue}{$28.25$}&$\pm2.98$ & \color{blue}{$26.34$}&$\pm2.54$ \\

\midrule
\multirow{2}{*}{ImageNet-Subset} & AlexNet & $43.11$&$\pm3.17$ & $46.67$&$\pm2.07$ & $46.54$&$\pm2.88$ & \color{blue}{$48.54$}&$\pm2.68$ \\
& VGG-16 & $38.94$&$\pm5.39$ & 
{$27.13$}&$\pm4.37$ & $39.57$&$\pm8.13$ & $35.53$&$\pm5.97$ \\

\bottomrule
\end{tabular}
\label{tab:novl_eval}
\end{table}

We evaluate the individual contribution of \ClassBalanced (CB-DW), Generator Ensemble (GE), and \SelectiveGeneration by progressively adding them to the baseline, DisGUIDE$^*$, which only differs from DisGUIDE by incrementing the replay iteration from 3 to 12 to match \ours' for a fair comparison. 
We report the averaged extraction accuracy with a 95\% confidence interval in~\cref{tab:novl_eval}. Column \emph{DisGUIDE$^*$$_\text{+CB-DW}$} represents 
DisGUIDE$^*$ with CB-DW. 
Other columns define similar ablations of our technique or our full approach. 
We run each experiment 9 times and evaluate the statistical significance using independent t-tests against  DisGUIDE$^*$.

With the inclusion of more components over the baseline method, there is an increase in the number of settings showing statistically significant improvements over the baseline. Column \emph{DisGUIDE$^*$$_\text{+CB-DW}$}   shows that adding \ClassBalanced results in four statistically significant improvements over the baseline while only adversely affecting one extraction result. Likewise, Column \emph{\emph{DisGUIDE$^*$$_\text{+CB-DW+GE}$} } shows that  Generator Ensembles with \ClassBalanced further increase the improved cases to 8 statistically significant improvements. Finally, our comprehensive approach, \ours, achieves statistically significant improvement in 10 out of 13 extraction settings, with only one setting where the performance is reduced, demonstrating the effectiveness of Selective Query.

\smallskip\noindent\fbox{
\centering
\parbox{0.95\linewidth}{
\textbf{Summary}: Each of the three \ours components, i.e., \ClassBalanced, Generator Ensemble, and \SelectiveGeneration,  improves model extraction accuracy.}}

\subsection{GE's Impact on Computational Cost}
\begin{table}[t]
\centering
\scriptsize
\caption{Evaluation of the impact Generator Ensemble Size on computation cost. Extraction result performed by \emph{DisGUIDE$^*$$_\text{+GE}$} following limited-budget setting on the AlexNet victim trained on CIFAR-10. We run each experiment setting 10 times.}
\vspace*{-\baselineskip}
\begin{tabular}{l@{\hspace{15pt}}c@{\hspace{5pt}}c@{\hspace{5pt}}c}
\toprule
\multirow{2}{*}{\textbf{Metric}} & \multicolumn{3}{c}{\textbf{Generator Ensemble Size}} \\
\cmidrule(lr){2-4}
 & \textbf{1} & \textbf{4} & \textbf{8} \\ \midrule
Accuracy (\%) & \hspace{2pt} 21.80$\pm$1.96 & \hspace{2pt} 28.56$\pm$2.79 & \hspace{2pt} 28.01$\pm$2.38 \\
Time (seconds) & 246.89$\pm$3.26 & 240.71$\pm$1.10 & 249.67$\pm$4.93 \\ \bottomrule
\end{tabular}
\vspace*{-\baselineskip}
\label{tab:ge_time_comparison}
\end{table}

We show in \cref{tab:ge_time_comparison} that under a fixed generator training iteration, the increase in Generator Ensemble size does not lead to a noticeable increase of our implementation's runtime. Increasing the generator size from 1 to 4 results in a 6.76\% increase in final accuracy, yet the time required for extraction remains approximately the same. In fact, the time is reduced by 6 seconds, which we believe is due to system randomness. This evaluation was performed on an NVIDIA GeForce RTX 2080 Ti with 11 GB of memory and an Intel(R) Xeon(R) Gold 5120 CPU.
 \section{Conclusion, Limitations, and Future Work}
We propose a data-free model extraction approach, \ours, which utilizes a combination of generator ensemble, \classbalanced, and \selectivegeneration. This approach enhances the accuracy and efficiency of model extraction results. In addition, \ours works in the more strict class-agnostic setting across seven different datasets and six \victim model architectures.

One limitation is that there is no easy way to tune hyperparameters in a DFME environment. \ours tries to use a robust set of parameters between experiment settings, varying only model learning rate between clone architectures and query budgets. However, the chosen settings may not generalize to untested datasets and architectures. Coming up with ways for the attacker to dynamically verify chosen hyperparameters is a challenging problem that needs to be solved for real world data-free model extraction to work.

Machine learning is a very broad field and many types of problems exist. Both prior work and \ours have naturally focused on a set of image data tasks, in order to be able to compare with one another. It remains to be seen how well current DFME SOTAs generalize to extracting models trained on class imbalanced data and non-image data. 

In addition, the attackers may have some domain knowledge about the training data. One promising future work direction is to utilize such domain knowledge effectively to improve mode extraction accuracy and efficiency. 

 \section*{Acknowledgements}
This work has been partially supported by NSF 2006688 and a J.P. Morgan AI Faculty Research Award.

\newpage

\bibliography{egbib}
\bibliographystyle{splncs04}

\appendix
\crefalias{section}{appendix}

\section{Fidelity Results}
In the Data-Free setting, prior works generally focus on replicating model performance. In this section we show the performance of \ours in terms of fidelity, i.e. how well the extracted model matches the victim model, as opposed to the true test label.

\begin{figure}
\centering
\begin{subfigure}{0.49\textwidth}
    \includegraphics[width=\textwidth]{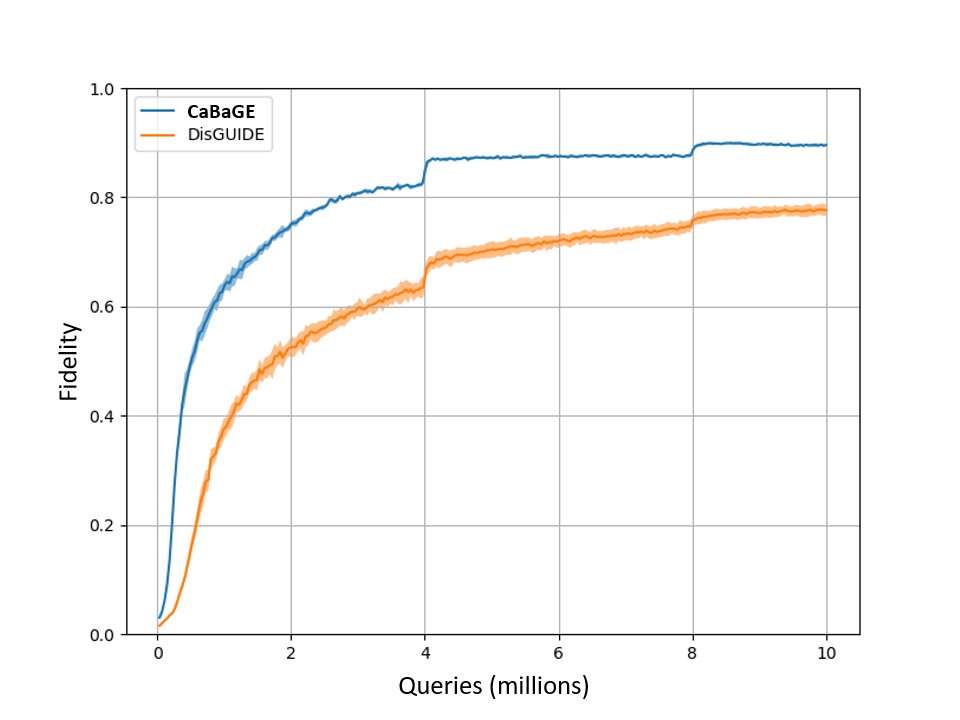}
    \caption{Fidelity extraction curve for CIFAR-100 SL. Orange is the prior SOTA for data-free model extraction in that setting. Blue is \ours}
\end{subfigure}
\begin{subfigure}{0.49\textwidth}
    \includegraphics[width=\textwidth]{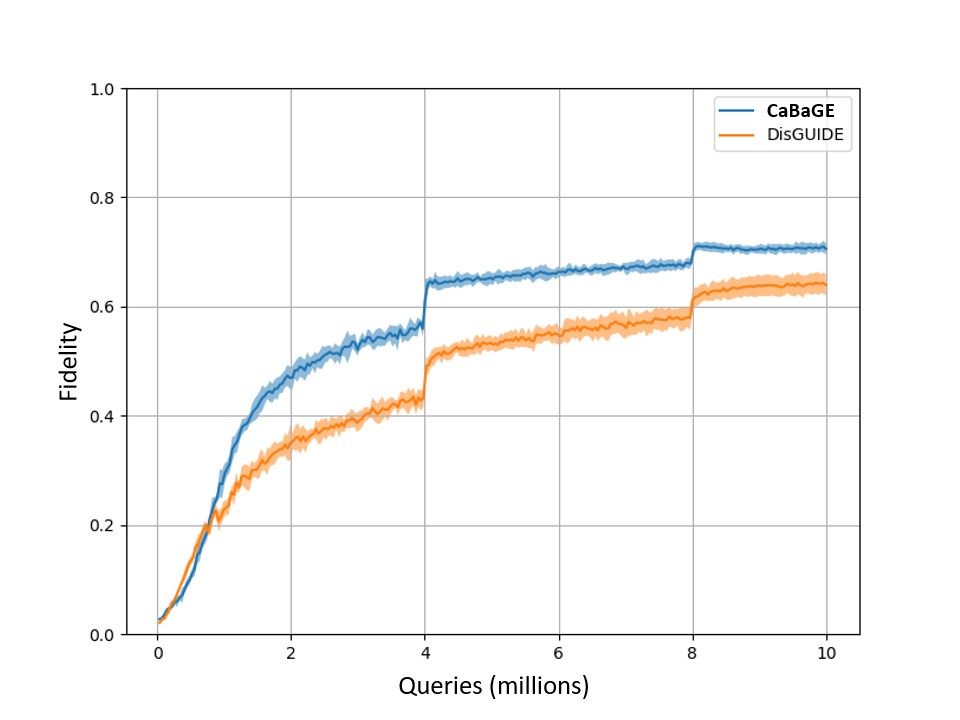}
    \caption{Fidelity extraction curve for CIFAR-100 HL. Orange is the prior SOTA for data-free model extraction in that setting. Blue is \ours}
\end{subfigure}

\begin{subfigure}{0.49\textwidth}
    \includegraphics[width=\textwidth]{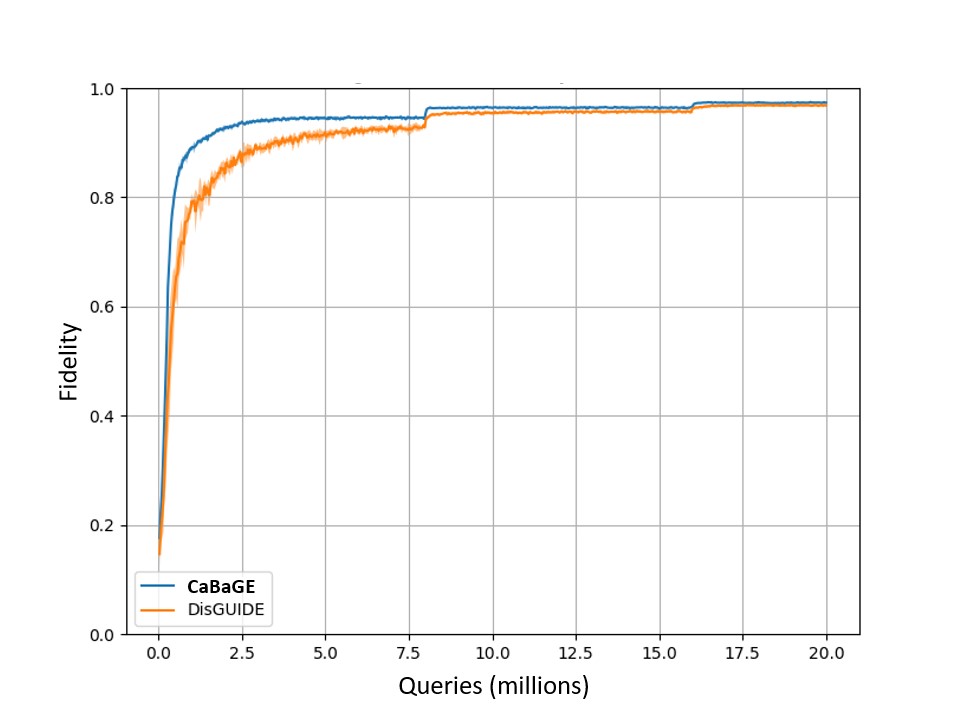}
    \caption{Fidelity extraction curve for CIFAR-10 SL. Orange is the prior SOTA for data-free model extraction in that setting. Blue is \ours}
\end{subfigure}
\begin{subfigure}{0.49\textwidth}
    \includegraphics[width=\textwidth]{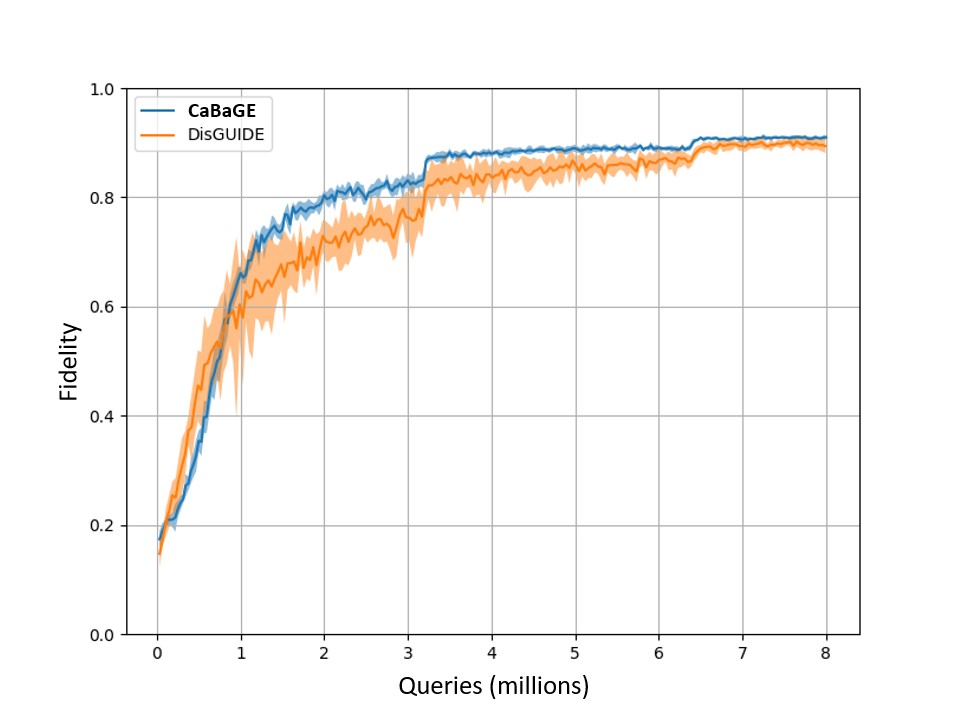}
    \caption{Fidelity extraction curve for CIFAR-10 HL. Orange is the prior SOTA for data-free model extraction in that setting. Blue is \ours}
\end{subfigure}
\caption{Fidelity extraction curves for CIFAR-10 and CIFAR-100}
\label{fig:fid_extract}
\end{figure}

\begin{table}[ht]
\scriptsize
\begin{center}
\caption{Final clone fidelity along with 95\% CI. Results from DisGUIDE are from reproduced runs with their codebase.}
\label{tab:method_comparison_fid}
\begin{tabular}{ll rr@{ }l rr@{ }l}
 \toprule
 \multirow{2}{*}{\textbf{Setting}} & \multirow{2}{*}{\textbf{Technique}} 
 & \multicolumn{3}{c}{\textbf{CIFAR-10}} & \multicolumn{3}{c}{\textbf{CIFAR-100}} \\
 
 \cmidrule(lr){3-5} \cmidrule(lr){6-8}
 
 &
 & \textbf{Victim}  & \multicolumn{2}{r}{\textbf{Clone (\%)}} 
 & \textbf{Victim}  & \multicolumn{2}{r}{\textbf{Clone (\%)}}\\
 
 \midrule
 \multirow{2}{*}{Soft Label} 
 & DisGUIDE  & 100 & 96.82 & $\pm$ 0.31 & 100 & 77.64 & $\pm$ 1.04 \\ 
 & \textbf{\ours}  & 100 & \textbf{97.32} & $\pm$ 0.09   & 100 & \textbf{89.56} & $\pm$ 0.19 \\ 
 \midrule
 \multirow{2}{*}{Hard Label} 
 & DisGUIDE  & 100 & 89.38 & $\pm$ 1.29  & 100 & 63.91 & $\pm$ 1.94 \\
 & \textbf{\ours}  & 100 & \textbf{90.98} & $\pm$ 0.41 & 100 & \textbf{70.54} & $\pm$ 1.02 \\

 \bottomrule
\end{tabular}
\end{center}
\end{table}

\cref{fig:fid_extract} compares the fidelity extraction curves of the prior SOTA with \ours. In the next section, we present the accuracy curves of the same runs in \cref{fig:acc_extract}. The final fidelity values for these runs can be found in \cref{tab:method_comparison_fid}. Subjectively the accuracy and fidelity results look to be inline with one another, with fidelity values being higher than accuracy.

\section{Accuracy Training Curves}

In section 5.2 we follow prior work in terms of quantifying the queries needed to reach the prior SOTA accuracy in different settings. Not limited in terms of space, we offer the reader accuracy training plots here.

\begin{figure}
\centering
\begin{subfigure}{0.49\textwidth}
    \includegraphics[width=\textwidth]{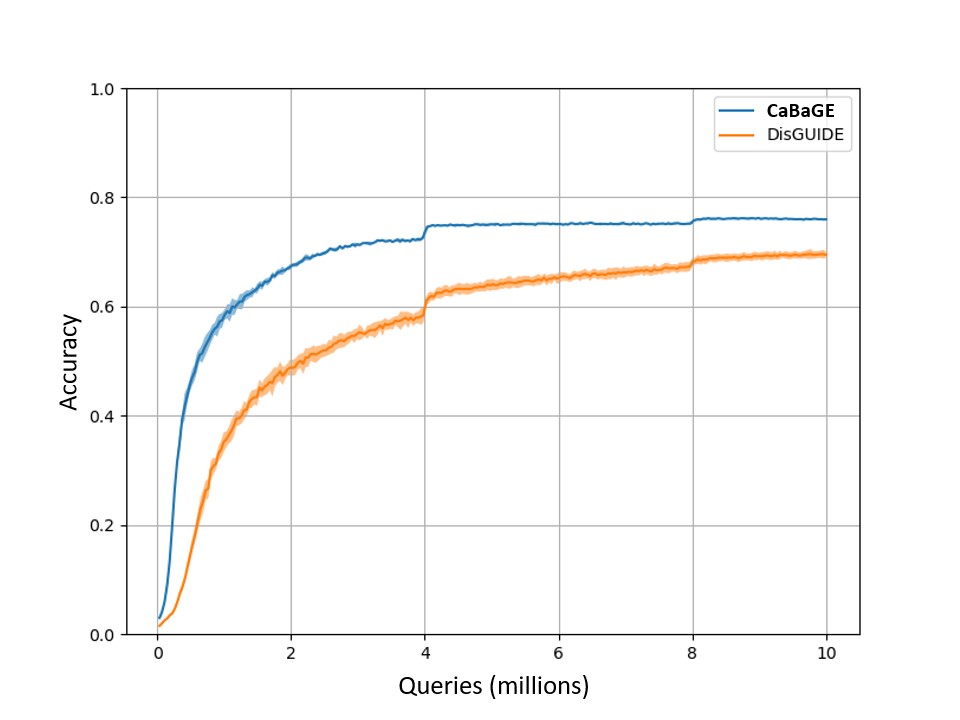}
    \caption{Accuracy training curve for CIFAR-100 SL extraction. Orange is the prior SOTA for data-free model extraction in that setting. Blue is \ours}
\end{subfigure}
\begin{subfigure}{0.49\textwidth}
    \includegraphics[width=\textwidth]{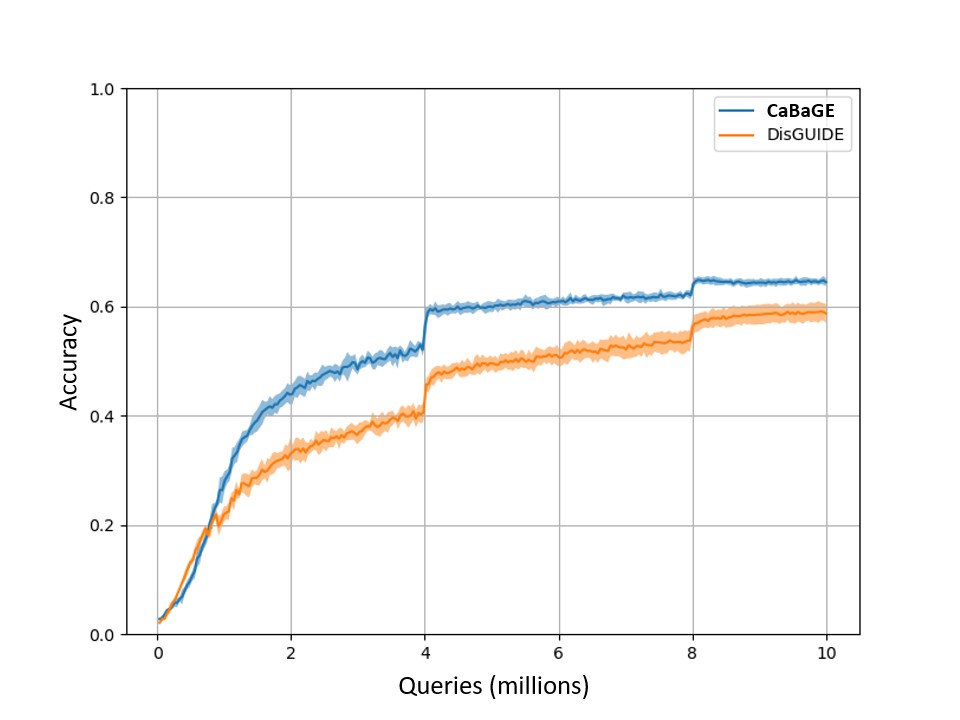}
    \caption{Accuracy training curve for CIFAR-100 HL extraction. Orange is the prior SOTA for data-free model extraction in that setting. Blue is \ours}
\end{subfigure}

\begin{subfigure}{0.49\textwidth}
    \includegraphics[width=\textwidth]{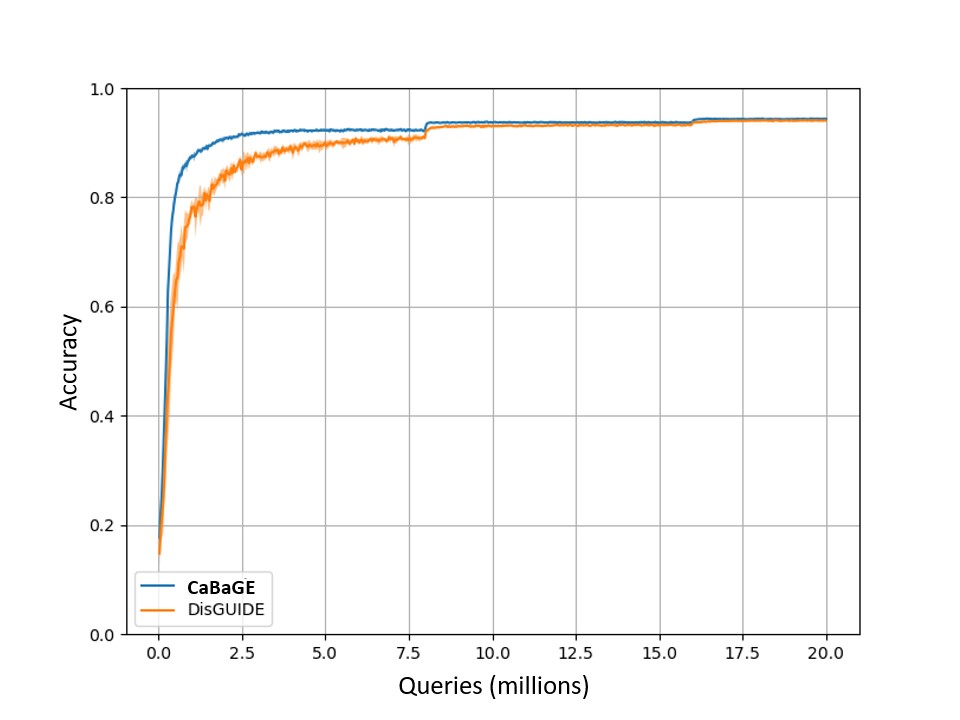}
    \caption{Accuracy training curve for CIFAR-10 SL extraction. Orange is the prior SOTA for data-free model extraction in that setting. Blue is \ours}
\end{subfigure}
\begin{subfigure}{0.49\textwidth}
    \includegraphics[width=\textwidth]{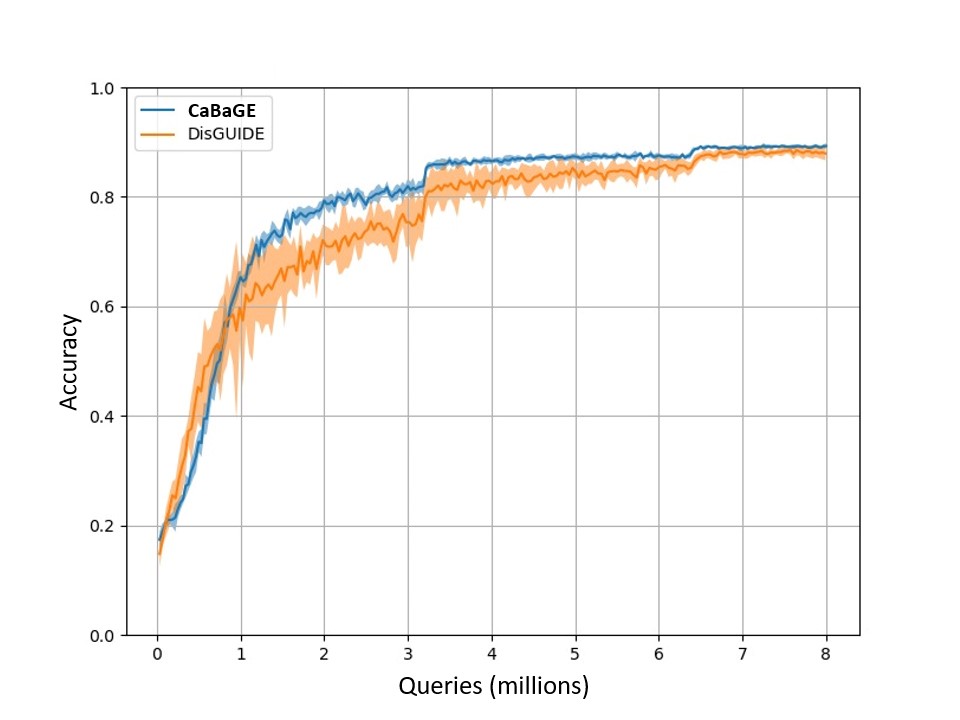}
    \caption{Accuracy training curve for CIFAR-10 HL extraction. Orange is the prior SoTA for data-free model extraction in that setting. Blue is \ours}
\end{subfigure}
\caption{Accuracy extraction curves for CIFAR-10 and CIFAR-100}
\label{fig:acc_extract}
\end{figure}

The plots in \cref{fig:acc_extract} are in the relaxed budget settings with the exact runs used to generate the \ours main results compared with reproduced runs of DisGUIDE with results inline with the numbers reported in the paper.

\section{Effect of Replay Iteration}
\label{sec:replay_iter}
DisGUIDE incorporated a circular buffer mechanism for retaining previously queried samples, in order to optimize the query budget utilization. The paper suggested an enhancement in performance with increased replay frequencies, a concept analogous to the frequency of memory bank updates in related literature~\cite{disguide}. Our analysis presents a more nuanced perspective on this claim. We evaluated the performance dynamics of DisGUIDE's replay against our proposed replay strategy across varying settings in two distinct model extraction scenarios: a multi-layer perceptron on the MNIST dataset and VGG-16 on the SVHN dataset. Both methods employed the same extraction technique derived from DisGUIDE, with the replay strategy being the sole variable.

\begin{figure}
    \centering
    \begin{subfigure}{0.49\textwidth}
        \includegraphics[width=\textwidth]{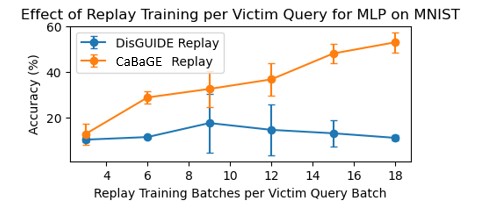}
        \caption{MLP extraction on MNIST}
        \label{fig:mnist_mlp}
    \end{subfigure}
    \begin{subfigure}{0.49\textwidth}
        \includegraphics[width=\textwidth]{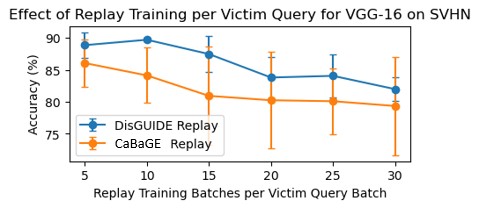}
        \caption{VGG-16 extraction on FMNIST}
        \label{fig:svhn-vgg}
    \end{subfigure}
    \caption{Performance comparison of DisGUIDE's replay and \ours, with different replay iterations used}
    \label{fig:replay_iter_compare_both}
\end{figure}

Figure \ref{fig:replay_iter_compare_both} demonstrates the performance trajectories of the two replays as a function of replay iterations, with the error bars indicating a 95\% confidence interval. The upper chart demonstrates that the performance of \ours's replay consistently exceeds that of DisGUIDE and exhibits a positive correlation with replay iterations when extracting the MLP victim on MNIST. Conversely, DisGUIDE's performance first increases then declines with additional iterations. In the lower chart, Nemesis replay marginally trails behind DisGUIDE's replay, and a uniform decrement in performance for both methods is observed as the replay iteration count escalates. These findings indicate that the influence of replay strategies is not uniform across datasets. It is imperative for future research in model extraction to leverage experience replay to scrutinize the differential impacts of replay modalities relative to the victim model and dataset characteristics.

\section{\Victim Training Details}

CIFAR-10 and CIFAR-100 ResNet-34 models were taken from the DFAD paper.

Due to the difficulty in acquiring the exact trained victim models from the IDEAL paper, new \victim models were trained for the purpose of running the \ours experiments. Aside from the DFAD models, all other victims were trained with the script provided in the codebase in run\_train\_teacher.sh

Models were trained via 240 epochs of stochastic gradient descent with a batch size of 250, momentum of 0.9, and an initial learning rate of 0.1. If the training run failed (model weights did not converge) the training was repeated with a smaller initial learning rate, until no such issues occurred. After every 40 epochs the learning rate was divided by 5. From within the training runs, models were selected to be close to the accuracies reported in the IDEAL paper, wherever possible.

\label{appendix:teacher_training}

\section{Reproducing IDEAL's result}
\label{appendix:Ideal}

As described in section 4.1, there is a discrepancy between the IDEAL codebase and paper. Specifically, IDEAL's extraction technique queries the victim many times with augmented versions of generated images. The transformation function from the codebase is given in the following code snippet:

\begin{lstlisting}
if not ("cifar" in dataset):
  self.transform = transforms.Compose(
    [
      transforms.RandomHorizontalFlip(),
      transforms.ToTensor(),
    ])
else:
  self.transform = transforms.Compose(
    [
      transforms.RandomCrop(32, padding=4),
      transforms.RandomHorizontalFlip(),
      transforms.ToTensor(),
    ])
\end{lstlisting}

For non-CIFAR datasets, the horizontally flipped versions of the inputs are used to query the \victim, but only 1 query is counted instead of 2. For CIFAR-10 and CIFAR-100, this is pushed further by also performing a random crop on generated images with a padding of 4. The random crop function adds padding to each side of an image and then randomly selects an section to match the dimensions specified by the first parameter. This means that vertically and horizontally there are 9 different possible outcomes (prepend 1-4 zeros, no change, or append 1-4 zeros). Combined with the horizontal flipping, this gives us $9 \times 9 \times 2 = 162$ different versions of each generated input, each of which the victim model may be queried with.

To remedy this issue, the transform is changed to the simply remove the transform, which is not mentioned in the paper. The code section in question after the fix is as follows:

\begin{lstlisting}
self.transform = transforms.Compose(
  [
    transforms.ToTensor()
  ])
\end{lstlisting}

\section{\SelectiveGeneration Specifics}
The \SelectiveGeneration pseudocode can be found in \textbf{Algorithm 1} \textit{\SelectiveGeneration}.

\begin{algorithm}
\label{pseudocode}
\small
\caption{\textit{\SelectiveGeneration 
}}\label{alg:selective generation}
    \begin{algorithmic}[]
        \State \textbf{Input}: $S = \{{S}_{i \in m}\}$,$m$ batches of generated images; $C(S)$, Clone ensemble's predictions on $S$; $K$, discovered classes; $N$, batch size.
        \State \textbf{Output}: $B$, selected data 
        \State $B \gets $ [ ]
        \State $N_k \gets \floor{\frac{N}{K}}$\hfill Expected number of samples per class
        \State$R \gets N - K\cdot N_k$ \hfill Number of missing samples
	\For {$k$ \textbf{in} $K$}
            \State $\{S\}_{k} \gets$ select images in $S$ with prediction $k$ in $C(S)$
            \State $\{S_{k}\}^{sorted}\gets sort(\{S\}_{k})$: Sort in descending order by corresponding value in $\sigma(C(S))$
            \If {$|\{S\}_{k}|\geq N_k$}
                \State add $\{{S_{k}}_j\}^{sorted}_{j\leq N_k}$ to $B$
            \Else
                \State add $\{S_{k}\}^{sorted}$ to $B$
                \State $R \gets R + N_k - |\{S\}_{k}|$
            \EndIf
	\EndFor 
        \If {$R > 0$}
            \State $S^{*sorted} = sort(\{S\setminus B\})$: Sort by decreasing value in $\sigma(C(S))$
            \State add $\{S^{*sorted}_i\}_{i\leq \floor{\frac{R}{2}}}$ to $B$
            \State add uniformly sampled samples from remaining to $B$
        \EndIf
        \State return $B$
    \end{algorithmic} 
\end{algorithm}

\section{Stability Against Hyper-parameters}
We study the impact of hyper-parameters, including the generator ensemble size $|G_e|$, and the generator training iterations per \victim query $g_{iter}$. 

\subsection{Effect of Generator Ensemble Size and Training Iterations}
\label{sec:abl_ge_gi}

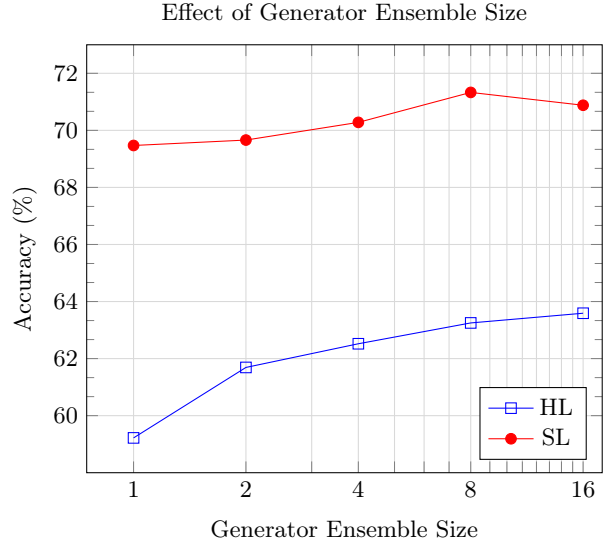
\begin{figure}
    \centering
    \begin{tikzpicture}[trim axis left, trim axis right]
        \begin{axis}[
            title={Effect of Generator Ensemble Size},
            xmode=log,
            xlabel={Generator Ensemble Size},
            ylabel={Accuracy (\%)},
            ylabel near ticks,
            xmin=0, xmax=18,
            ymin=58, ymax=73,
            xtick={1,2,3,4,5,6,7,8,9,10,11,12,13,14,15,16},
            xticklabels={1,2,,4,,,,8,,,,,,,,16},
            scaled x ticks=false,
            ytick={60,62,64,66,68,70,72},
            legend pos=south east,
            grid=major,
            minor tick num=2,
            major grid style={gray!30},
            minor grid style={gray!10},
        ]
        
        \addplot[
            color=blue,
            mark=square,
        ]
        coordinates {
            (1,59.22)
            (2,61.69)
            (4,62.52)
            (8,63.25)
            (16,63.59)
        };
        \addlegendentry{HL}
        
        \addplot[
            color=red,
            mark=*,
        ]
        coordinates {
            (1,69.47)
            (2,69.66)
            (4,70.28)
            (8,71.33)
            (16,70.88)
        };
        \addlegendentry{SL}

        \end{axis}
    \end{tikzpicture}
    \caption{Final accuracy comparison for ResNet34 extraction on CIFAR-100 under the relaxed-budget setting. Method tested is DisGUIDE with only the addition of a generator ensemble. Effect of varying the ensemble size.}
    \label{fig:gen_size}
\end{figure}

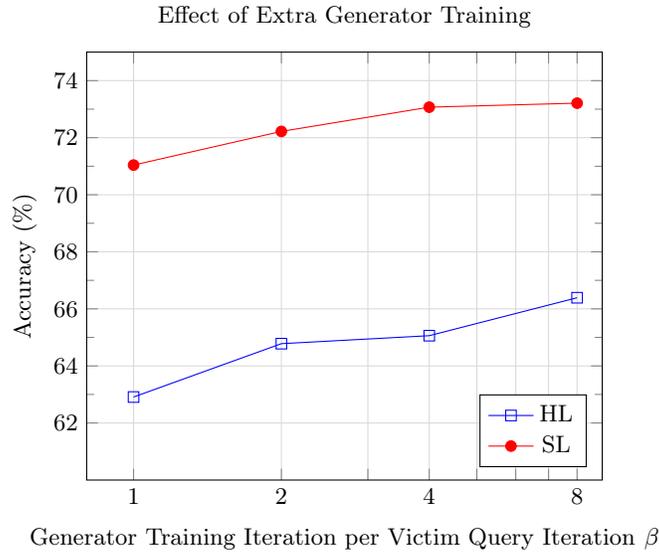
\begin{figure}
    \centering
    \begin{tikzpicture}[trim axis left, trim axis right]
        \begin{axis}[
            title={Effect of Extra Generator Training},
            xmode=log,
            xlabel={Generator Training Iteration per Victim Query Iteration $\beta$},
            ylabel={Accuracy (\%)},
            ylabel near ticks,
            xmin=0, xmax=9,
            ymin=60, ymax=75,
            xtick={1,2,3,4,5,6,7,8},
            xticklabels={1,2,,4,,,,8},
            ytick={62,64,66,68,70,72,74},
            legend pos=south east,
            grid=major,
            minor tick num=1,
            major grid style={gray!30},
            minor grid style={gray!10},
        ]
        
        \addplot[
            color=blue,
            mark=square,
        ]
        coordinates {
            (1,62.91)
            (2,64.78)
            (4,65.06)
            (8,66.39)
        };
        \addlegendentry{HL}
        
        \addplot[
            color=red,
            mark=*,
        ]
        coordinates {
            (1,71.04)
            (2,72.22)
            (4,73.07)
            (8,73.21)
        };
        \addlegendentry{SL}

        \end{axis}
    \end{tikzpicture}
    \caption{Final accuracy comparison for ResNet34 extraction on CIFAR-100 under the relaxed-query setting. Generator ensembles' size is fixed to 16.}
    \label{fig:gen-iter}
\end{figure}
We explored the impact of both \textbf{generator ensemble size} and \textbf{generator training iterations} on the CIFAR-100 dataset, and our experiments results are encapsulated in~\cref{fig:gen_size} and~\cref{fig:gen-iter} respectively. To ensure a fair comparison, our experimental setup solely applies the generator ensemble technique to DisGUIDE, excluding the integration of \SelectiveGeneration and \classbalanced. The configurations mirror the DisGUIDE setup for CIFAR-100 as detailed in~\cref{sufficient_qb_acc_comparison}~\cite{disguide}.

In experiments for evaluating the generator ensemble size, shown in~\cref{fig:gen_size}), for both \textit{SL Setting} and \textit{HL Setting}, the extracted model's final accuracy generally improves with an increase in ensemble size, peaking at a size of 8, which results in $3.48\%$ improvement over base DisGUIDE with a final accuracy of $66.39\%$ . However, expanding the ensemble from 8 to 16 members results in a marginal decline in performance.
         
The impact of \textit{generator training iterations per \victim query iteration} are shown in~\cref{fig:gen-iter}. In both the \textit{HL Setting} and the \textit{SL Setting}, A clear upward trend is observed with increased training iterations. The increments in accuracy for each increasing generator training iteration ($\beta$) are as follows. For \textbf{HL}: an increase of 1.87\% from $\beta=1$ to $\beta=2$, 0.28\% from $\beta=2$ to $\beta=4$, and 1.33\% from $\beta=4$ to $\beta=8$. For \textbf{SL}: an increase of 1.18\% from $\beta=1$ to $\beta=2$, 0.85\% from $\beta=2$ to $\beta=4$, and 0.14\% from $\beta=4$ to $\beta=8$. The result suggests that more generator training leads to enhanced performance in both HL and SL settings. However, similar to the generator size, we can observe the diminishing returns and it is possible that the increased generator training iterations' positive effect on model performance will be reversed beyond a specific threshold.

\subsection{Comparison Against Improved Baseline}
\begin{table}
    \centering
    \scriptsize 
    \caption{Independent t-test comparing \ours (using \classbalanced ) and DisGUIDE (using original replay) under 12 replay iterations in the setting of various configurations. There are 9 runs for all experiments. accuracy is shown along with 95\% confidence interval.}
    \begin{tabular}{l@{\hspace{10pt}}l@{\hspace{10pt}}r@{}l@{\hspace{10pt}} r@{}l@{\hspace{10pt}} l@{\hspace{10pt}}l}
    \toprule
    \textbf{Dataset} & \textbf{Model} & \multicolumn{2}{c}{\hspace{-7pt}\textbf{DisGUIDE$^*$}} & \multicolumn{2}{c}{\hspace{-10pt}\textbf{\ours}} & \textbf{\hspace{4pt}$p$ val} & \textbf{Better} \\
    \midrule
& MLP & $11.35$&$\pm0.00$ & $54.94$&$\pm3.58$& $p$=0.000 & \ours \\
MNIST & LeNet & $94.20$&$\pm0.55$ & $95.32$&$\pm0.63$& $p$=0.011 & \ours \\
& AlexNet & $74.10$&$\pm2.11$ & $86.08$&$\pm0.94$& $p$=0.000 & \ours \\

\midrule
& MLP & $42.28$&$\pm4.40$ & $74.45$&$\pm3.28$& $p$=0.000 & \ours \\
FMNIST & LeNet & $69.12$&$\pm1.12$ & $68.08$&$\pm1.50$& $p$=0.246 & N/A \\
& AlexNet & $74.66$&$\pm2.09$ & $79.63$&$\pm1.48$& $p$=0.001 & \ours \\

\midrule
& VGG-16 & $88.37$&$\pm0.87$ & $84.14$&$\pm2.09$& $p$=0.002 & DisGUIDE \\
SVHN & ResNet-18 & $76.05$&$\pm1.81$ & $78.70$&$\pm0.98$& $p$=0.016 & \ours \\
& AlexNet & $64.01$&$\pm2.40$ & $74.46$&$\pm2.19$& $p$=0.000 & \ours \\

\midrule
\multirow{2}{*}{CIFAR-10} & AlexNet & $26.25$&$\pm3.45$ & $33.23$&$\pm1.04$& $p$=0.002 & \ours \\
& ResNet-34 & $22.04$&$\pm2.64$ & $26.34$&$\pm2.54$& $p$=0.021 & \ours \\

\midrule
\multirow{2}{*}{ImageNet$_{12}$} & AlexNet & $43.11$&$\pm3.17$ & $48.54$&$\pm2.68$& $p$=0.012 & \ours \\
& VGG-16 & $38.94$&$\pm5.39$ & $35.53$&$\pm5.97$& $p$=0.370 & N/A \\

    \bottomrule
    \end{tabular}
    \label{tab:disguidevsDGP-rep12}
    \end{table}
In~\cref{tab:disguidevsDGP-rep12}, we present a performance comparison between DisGUIDE and \ours method while eliminating the effect of replay iterations. The experimental configurations remain consistent with those in~\cref{tab:qblimited_accuracy}, except that we increased DisGUIDE's replay iteration from 3 to 12, and denote this method as DisGUIDE$^*$. Based on our findings in~\cref{sec:replay_iter}, we have observed that increasing the replay iterations can lead to a boost in the final accuracy of the extracted model, although the gains diminish over time. To provide a fair comparison between the two methods, we chose to set DisGUIDE's replay iterations equal to \ours's, and we run all experiments 9 times. The results are shown as mean values of extracted models' accuracies, along with standard deviations. Statistical significance is determined using independent t-tests, and the $p$ value for each test is recorded (Column $p$ val) with three decimal places.

\ours demonstrates superior performance in most of the configurations. Setting a significance level at $\alpha = 0.05$, most results demonstrate statistical significance. Specifically, \ours outperforms DisGUIDE$^*$ in 10 out of the 13 configurations. \ours is only outperformed by DisGUIDE$^*$ while extracting a VGG-16 model on the SVHN dataset.

\section{Remaining Ablation on Larger Datasets}
\begin{table}
    \centering
    \scriptsize
    \caption{Final Accuracy of \ours and ablation settings shown with 95\% confidence interval. Independent t-tests are used to determine improvement over the baseline DisGUIDE$^*$, where statistically significant improvements are highlighted in \color{blue}{Blue}\color{black}.
}
    
    \begin{tabular}{l@{\hspace{14pt}}l@{\hspace{14pt}}rl@{\hspace{14pt}}rl@{\hspace{14pt}} 
rl@{\hspace{14pt}}rl}

\toprule

\multirow{2}{*}{\textbf{Dataset}} & \multirow{2}{*}{\textbf{Model}} & \multicolumn{2}{l}{\hspace{-3pt}\textbf{DisGUIDE$^*$}} & \multicolumn{2}{l}{\hspace{-3pt}\textbf{DisGUIDE$^*$}} & \multicolumn{2}{l}{\hspace{-3pt}\textbf{DisGUIDE$^*$}} & \multicolumn{2}{c}{\multirow{2}{*}{\textbf{\ours}}} \\

 & & \multicolumn{2}{c}{} & \multicolumn{2}{l}{\hspace{1pt}\textbf{$_\text{+CB-DW}$}} & \multicolumn{2}{l}{\hspace{-6pt}\textbf{$_\text{+CB-DW+GE}$}} & \multicolumn{2}{c}{} \\

    \midrule
\multirow{2}{*}{CIFAR-100} & AlexNet & $25.22$&$\pm1.98$ & $25.74$&$\pm2.04$ & \color{blue}{$34.68$}&$\pm1.30$ & \color{blue}{$33.09$}&$\pm0.54$ \\
 & ResNet-34 & $34.71$&$\pm1.28$ & $38.99$&$\pm1.69$ & \color{blue}{$43.00$}&$\pm0.69$ & \color{blue}{$43.75$}&$\pm1.75$ \\

\midrule
\multirow{2}{*}{Tiny ImageNet} & ResNet-34 & $13.70$&$\pm0.97$ & $12.15$&$\pm0.60$ & \color{blue}{$18.02$}&$\pm1.13$ & $15.36$&$\pm0.64$ \\
& VGG-16 & $9.68$&$\pm1.19$ & $9.98$&$\pm0.73$ & \color{blue}{$13.43$}&$\pm1.10$ & $11.98$&$\pm0.64$ \\

    \bottomrule
    \end{tabular}

    \label{tab:dgp_minus_extra}
    \end{table}
We show the additional experiments on CIFAR-100 and Tiny-Imagenet for our ablation on the individual contribution of our novel components in~\cref{tab:dgp_minus_extra}. All settings are the same as Sec. 5.2 in our paper, except all experiments are run 3 times instead of 9 times on this table due to time constraints. We progressively add our novel components to the baseline, DisGUIDE$^*$, and present the ablations for DisGUIDE$^*$, \emph{DisGUIDE$^*$$_\text{+CB-DW}$}, \emph{DisGUIDE$^*$$_\text{+CB-DW+GE}$} and \ours.

When including \classbalanced, the performance does not differ from the baseline much, which indicates that \classbalanced has a similar performance compares to the circular buffer replay used by DisGUIDE$^*$. Conversely, the integration of GE consistently demonstrates statistically superior results compared to the baseline on both CIFAR-100 and Tiny-ImageNet, highlighting the robust benefits of GE. Regarding our comprehensive method, \ours, which also includes SQ, there are two settings where it outperforms the baseline. While this may appear less effective compared to \emph{DisGUIDE$^*$$_\text{+CB-DW+GE}$}, it is important to note that the inclusion of SQ generally leads to a decrease in uncertainty levels, suggesting that SQ enhances the stability of the extraction results.

\end{document}